**Alexander L. Rudolph** is a professor of physics and astronomy at California State Polytechnic University in Pomona and the director of the Cal-Bridge program.

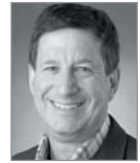

# Cal-Bridge

## Creating pathways to the PhD for underrepresented students in physics and astronomy

Alexander L. Rudolph

**The Cal-Bridge program connects promising juniors and seniors from underrepresented groups with STEM faculty mentors to help smooth the transition from undergraduate to graduate programs.**

The challenge of creating equal representation in STEM (science, technology, engineering, and math) is a longstanding one that has resisted improvement. The problem is especially stark at the PhD level. The number of underrepresented minorities (URMs), comprising the groups Latinx or Hispanic, Black or African American, and Native American, receiving STEM PhDs has remained around 14%, even though those groups make up more than 30% of the US population.[1] The problem is even more acute in physics and astronomy, where the percentage of PhDs awarded to URMs in 2016 was only 6% of the total. Women are also underrepresented, making up only 20% of PhDs in physics and astronomy, lagging even the rest of the STEM fields (see figure 1).

# CAL-BRIDGE

Attempts to address the lack of representation have had limited success. However, some recent programs are beginning to make progress. One of them is the Cal-Bridge program, a partnership between 9 University of California (UC), 16 California State University (CSU), and more than 30 community college campuses in California. The mission of Cal-Bridge is to increase the numbers of traditionally underrepresented groups in PhD programs in physics, astronomy, and closely related fields. More than 160 physics and astronomy faculty from the three systems participate in the program. The Cal-Bridge model has the potential to improve representation and inclusion in STEM PhD programs.

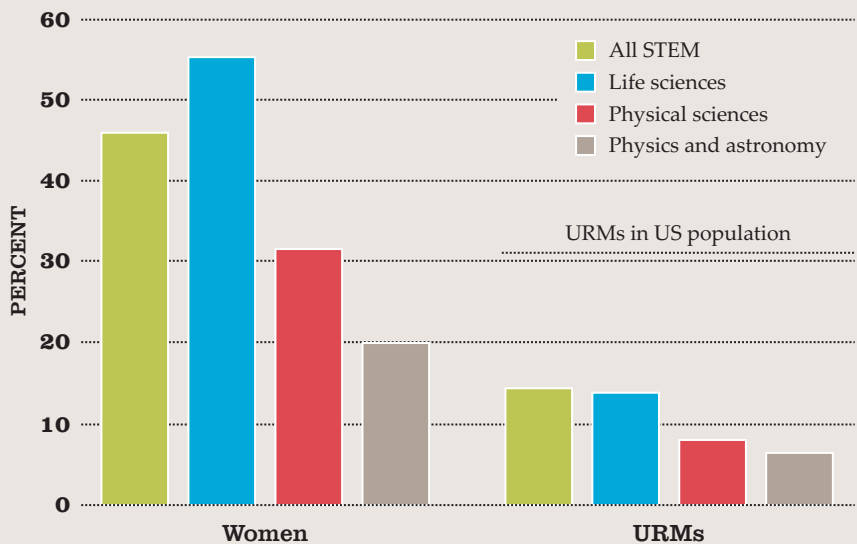

**FIGURE 1. PERCENTAGE OF STEM PHDS AWARDED IN 2016** to women and to members of underrepresented minorities (URMs). Numbers of PhDs awarded to members of both groups (for example, women of color) are even lower and are not reported separately by NSF. (Adapted from ref. 1.)

## Problems and progress

Reducing inequities at all educational levels is crucial both for creating equal opportunities and for ensuring the future health of the US scientific community. In the National Academies report *Expanding Underrepresented Minority Participation*, the authors note that "the S&E [science & engineering] workforce is large and fast-growing: more than 5 million strong and projected by the U.S. Bureau of Labor Statistics to grow faster than any other sector in coming years. This growth rate provides an opportunity to draw on new sources of talent, including underrepresented minorities, to make our S&E workforce as robust and dynamic as possible."[2]

Students with STEM degrees have a wide range of careers open to them (see the article in this issue by Anne Marie Porter and Susan White, page 32), and unemployment rates decrease rapidly with increasing education, from 5.3% for high school graduates, to 2.5% for those with a bachelor's degree, and even lower for those with advanced degrees.[3] About 12% of PhDs in STEM eventually attain faculty jobs,[4] where they become the teachers and role models for the next generation of college students. Although that percentage is relatively small, diversifying the faculty in physics and astronomy is an important goal. Studies have consistently shown that a lack of faculty role models dissuades students from underrepresented groups from choosing a STEM major.[5] In physics and astronomy, only 16% of faculty are women and 5% are URMs.[6] Inequalities in the professoriate and workforce are thus intertwined.

The percentage of URMs and women in physics and astronomy PhD programs has been slow to change, and the problem is even more acute for those with multiple underrepresented identities, such as women of color, even as the physics community recognizes those percentages as a problem (see the article by Jennifer Blue, Adrienne Traxler, and Ximena Cid, PHYSICS TODAY, March 2018, page 40). But some recent efforts are beginning to bear fruit. For example, there has been a small increase in the number of Hispanic PhDs in those fields in the past few years,[1] possibly due to the growth of PhD bridge programs such as Cal-Bridge, the Fisk–Vanderbilt Master's-to-PhD Bridge Program,[7,8] Columbia University's Bridge to the PhD Program, and the American Physical Society (APS) Bridge Program.

Other than Cal-Bridge, those programs are all based on a postbaccalaureate model. The oldest, Fisk–Vanderbilt, is an innovative partnership between Fisk University, a prominent historically black institution, and Vanderbilt University, a top research university located only two miles away. The program focuses on the master's degree as a key pathway to the PhD for URM students. Minority students are approximately 50% more likely to seek a master's degree on their way to a PhD than are nonminority students.[9] The Fisk–Vanderbilt program currently grants 10 times the national average of URM PhDs in astronomy and 5 times the national average in physics.

The APS Bridge Program recruits candidates nationally and matches them with dozens of vetted graduate programs, mostly PhDs, in physics. Two-thirds of applicants to the APS program had not been admitted to a PhD program; the other third had not applied at all, often due to perceived deficiencies in their GPA or physics GRE score. As of 2019 the program had placed more than 200 applicants into bridge programs or partner sites, including 40 students in 2016 alone, the last year they reported (see the article by Ted Hodapp and Kathryne Woodle, PHYSICS TODAY, February 2017, page 50).

## Cal-Bridge: A different model

The bridge programs described above serve postbaccalaureate students who did not transition directly to a PhD program after receiving a bachelor's degree. The Cal-Bridge model offers a different approach: ensuring adequate preparation and broadening faculty attitudes in the PhD admissions process *before* students graduate with a bachelor's degree.

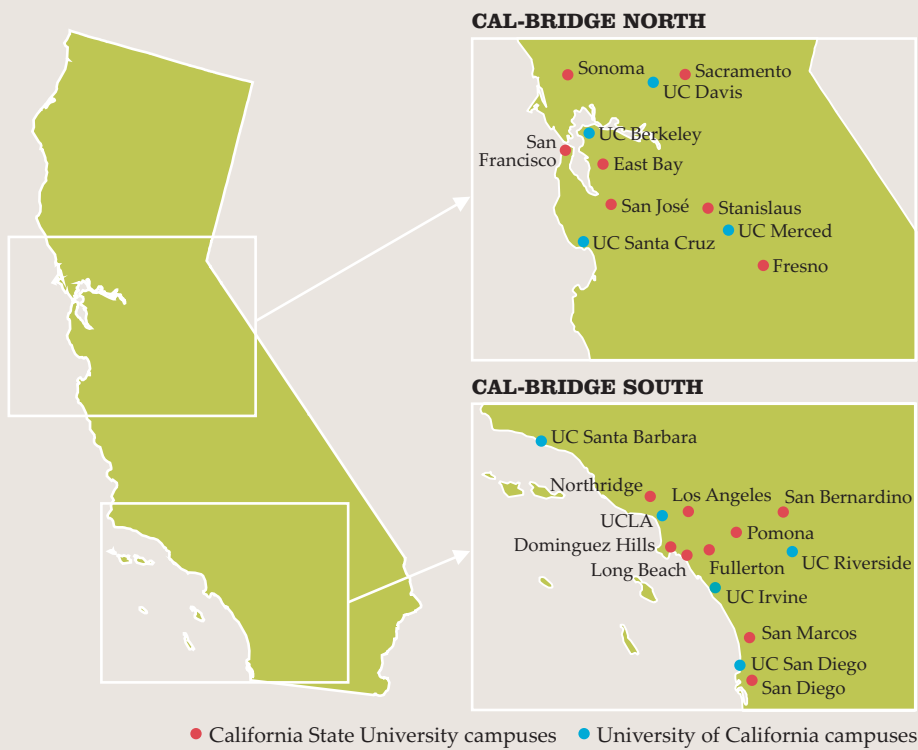

**FIGURE 2. LOCATIONS OF CAL-BRIDGE CAMPUSES.**

To achieve that end, the Cal-Bridge program recruits and supports students entering their last two years of undergraduate studies. In 2015 almost 70% of undergraduate URM students interested in STEM did not complete their STEM degree, and few proceeded to pursue a PhD.[10] Given the high attrition rate, it is critical to support students as early as possible. Our hope is to eliminate the necessity for the detour many underrepresented students take by obtaining master's training at one university before obtaining a PhD at another institution. We also hope to identify and recruit students who never even make it to the stage of applying to programs like Fisk–Vanderbilt and APS Bridge, let alone a PhD program, because they struggled early in their undergraduate career or lacked awareness of the PhD as a possibility.

The Cal-Bridge program is divided into two parallel subprograms, Cal-Bridge South and Cal-Bridge North (see figure 2). Recruiting takes place at the 16 participating CSU campuses, plus more than 30 community colleges that are primary feeders for transfer students to CSUs. Faculty mentors at the CSUs and one or more liaisons at each community college are the primary program recruiters. They are responsible for identifying, cultivating, and mentoring potential applicants. That form of active recruiting is critical to increasing diversity; we have found that many students from underrepresented groups lack basic knowledge about a PhD as a path to follow or do not feel welcome in our fields unless specifically encouraged.

Cal-Bridge selects scholars using a model similar to Fisk–Vanderbilt, which is based on social science research and employs specific criteria and practices.[5] Applications are submitted online and include three essay questions designed to elicit information about the motivation and capabilities of the applicants. Each region has its own steering committee, which consists of UC, CSU, and community college faculty. The steering committees for each region review the initial applicant pool. A group of finalists is selected for in-depth 30-minute interviews via video conference with two steering committee members, one from a CSU or community college and one from the UC system. The steering committees then meet to select the Cal-Bridge scholars for their region based on the criteria in the interview protocol.

During the process, steering committee members review the applicant's personal essays and letters of recommendation to assess the student's work ethic, initiative, focus, and perseverance; consider their academic performance in math and physics courses; and evaluate their community service, leadership, and outreach activities as indicators of motivation and long-term goals. Committee members also focus on the student's academic potential as evidenced by performance in individual courses and improvement over time, and take into account situations where a student might have an uneven record due to external demands like work, family, or psychosocial stressors.

## Program structure

The National Academies report *Expanding Underrepresented Minority Participation* highlights two key priorities for broadening participation in the STEM workforce. To address the first priority, undergraduate retention and completion, the report proposes that higher education institutions provide "strong academic, social, and financial support . . . along with programs that simultaneously integrate academic, social, and professional development."[11] For the second priority, transition to graduate study, it encourages programs that support the transition from undergraduate to graduate education and provide support in graduate programs.

The design and implementation of Cal-Bridge reflects those priorities and solutions by building the program on four pillars: financial support, mentoring, cohort building and professional development, and research experiences. In addition, faculty at both the scholars' home institutions and at the institutions where the scholars hope to matriculate to obtain their PhD are active participants. We next describe the resources for scholars and participating faculty under each pillar.

# CAL-BRIDGE

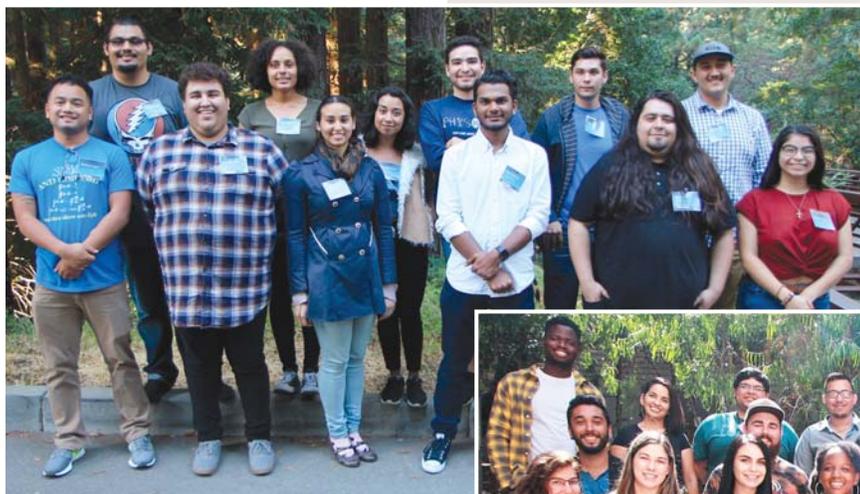

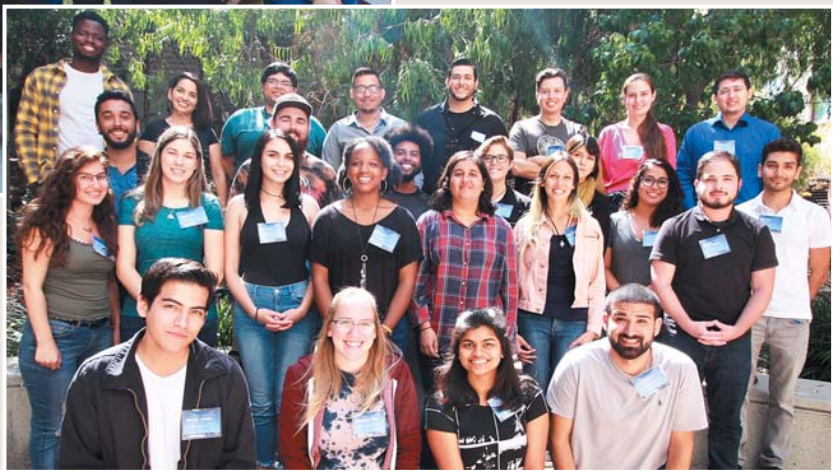

**FIGURE 3. CAL-BRIDGE NORTH (LEFT) AND SOUTH (BELOW) SCHOLARS** at fall 2018 orientation. (Courtesy of the Cal-Bridge program.)

## Financial support

Most applicants to Cal-Bridge have demonstrated financial need beyond the aid they already receive from the state and federal governments. Based on 2016–17 data, 80% of CSU students receive financial aid, despite the relatively low cost of attendance: In-state resident tuition and fees are about $7000, and, depending on housing arrangements (on campus versus off campus), the full cost of attendance ranges from $15 000 to $25 000, including room, board, books, and so on. Of the CSU students receiving aid, 61% received Pell Grants. Average parental income of students receiving aid in the CSU system is under $45 000.

Cal-Bridge scholars are given two years of need-based scholarship support, up to $10 000 per year, at their CSU campus to supplement any grants or scholarships they receive. The average need-based grant of Cal-Bridge scholars has been $9400. Most CSU students work to pay for their education and to help provide financial support to their families. Many scholars have been working 20 to 30 hours per week during their first two years of college, and some even hold full-time jobs that take valuable time away from their studies and from research and mentoring opportunities. As a condition of participation, Cal-Bridge scholars are limited to working fewer than 10 hours per week; thus financial support from Cal-Bridge allows students to remain engaged in their classes and in program activities.

## Mentoring

Following the Fisk–Vanderbilt model, each scholar is formally assigned two mentors: a CSU faculty member at their home institution and a UC faculty member. Mentoring programs have been shown to improve persistence, student performance, and academic self-esteem.[12,13] Joint mentoring is the best way to track student progress and to ensure scholars' readiness for PhD-level graduate work. Mentoring takes place via twice-monthly meetings between each scholar and their two mentors. The CSU faculty mentor gives academic advice on course selection and study habits, guides students toward research opportunities, and helps them apply to graduate school. The UC mentor performs similar functions and is also especially qualified to provide guidance toward readiness for graduate school.

The participation of a UC faculty mentor is a novel and critical piece of the Cal-Bridge model. Exposure to regular advice and validation from a faculty member at a PhD program gives added weight to the encouragement and guidance students receive. The UC mentor also advises the scholars on what is expected in a graduate school application and guides them through the application process.

Mentors also obtain monthly feedback from each scholar's instructors in order to track academic progress and catch problems early. That system allows for intervention when necessary; for example, mentors might help scholars work on their study habits or connect them with UC graduate student tutors, paid for by the program. To ensure that Cal-Bridge mentors implement research-based best practices, mentoring experts provide training workshops for Cal-Bridge faculty participants. Experienced faculty mentors also act as "near-peer" mentors to newer Cal-Bridge faculty.

Cal-Bridge also facilitates regular meetings among the faculty and students from different campuses and systems. Those meetings enable an exchange of information about how undergraduates should train and prepare for graduate school. Students and CSU and community college faculty members gain insight into graduate admissions decisions with the help of UC faculty who sit on PhD admissions committees.

Simultaneously, the UC faculty members gain reciprocal insight into the lives of CSU students, which greatly increases the faculty's awareness of both the challenges those students face and the strengths they can bring to a graduate program, strengths that may not be reflected in a paper application. When those UC faculty members evaluate applications to their PhD programs, their experience with Cal-Bridge may broaden the network of institutions and faculty recommenders that they trust to endorse qualified applicants. That familiarity can minimize the perceived risks of admitting students from institutions whose reputations UC faculty are less familiar with.[14] Identifying changes to the recruiting and admissions process that smooth the transition of underrepresented students into graduate school and building an institutional apparatus to support those changes are primary long-term goals of the Cal-Bridge program.

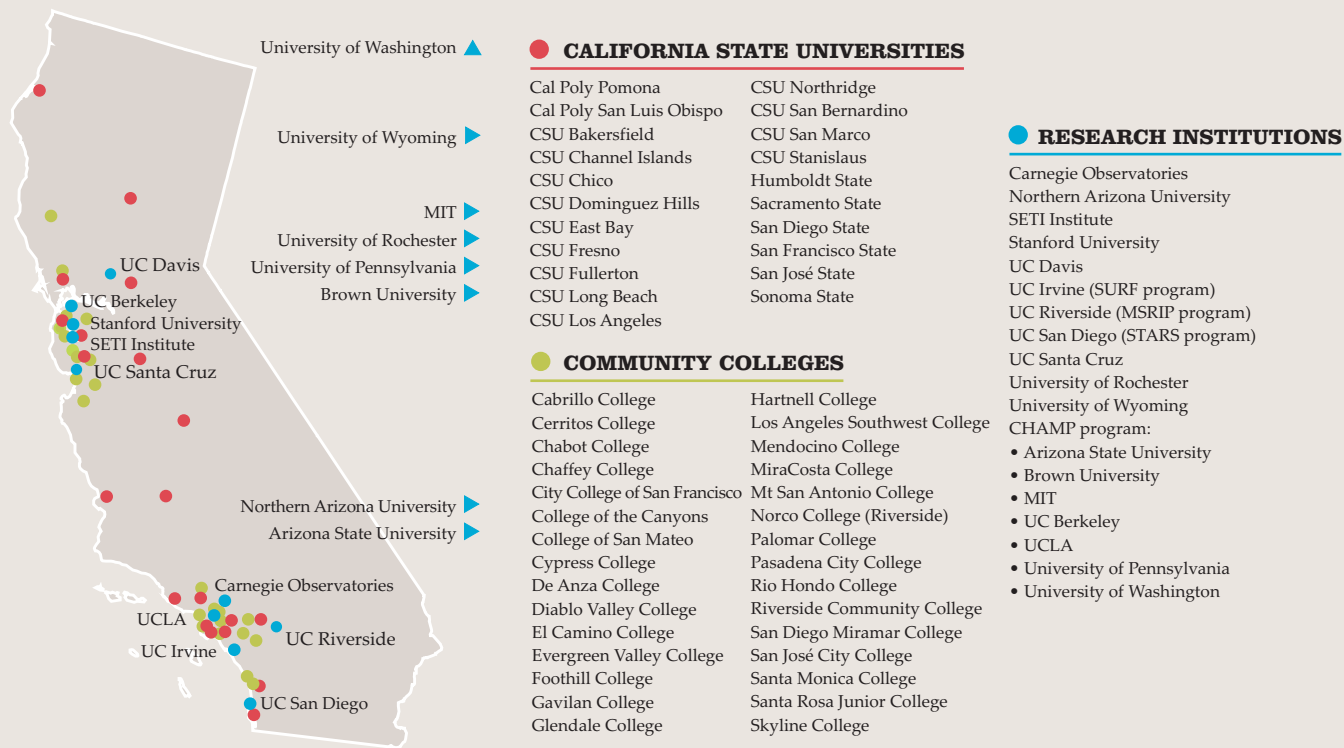

**FIGURE 4. LOCATIONS OF COLLEGES AND UNIVERSITIES IN THE CAMPARE NETWORK.** CAMPARE, a sister program to Cal-Bridge, matches promising undergraduates with summer research opportunities in STEM fields. The CAMPARE HERA Astronomy Minority Partnership (CHAMP) is a subprogram of CAMPARE that partners with the Hydrogen Epoch of Reionization Array (HERA) project.

## Cohort-building and professional development

To foster holistic support for young scholars, Cal-Bridge also holds regular monthly cohort-building, skill-building, and professional development workshops. Most are combined with a visit to a UC campus. The geographic compactness of the Cal-Bridge schools in each region, as seen in figure 2, makes it feasible for Cal-Bridge scholars to attend in-person workshops. That attendance in turn supports the creation of a peer cohort among Cal-Bridge scholars, an essential program element that increases retention.[15]

In the new scholar orientation, newly selected students (see figure 3) are introduced to the program and presented with the Cal-Bridge scholar contract, which outlines the obligations of the program and establishes clear expectations. Returning scholars work on developing a list of graduate schools to which they plan to apply and on refining their graduate school admissions essays. Other workshops are held throughout the year; topics include Python programming and cultivating a growth mindset.

Every spring, two workshops are held on graduate admissions essay writing. Scholars in their junior year learn the best practices for writing such essays at the first workshop and revise their drafts at the second. Over the summer their essays are reviewed by their peers, then by an eight-member UC faculty committee, and finally by their mentors before they summit them to graduate programs in the fall. All senior-year scholars are also required to apply for an NSF Graduate Research Fellowship. Seven scholars, representing 26% of Cal-Bridge scholars who applied, have received a fellowship in the past four years.

## Research experiences

Numerous studies have documented the benefits of undergraduate research in catalyzing interest in graduate education.[16,17] Cal-Bridge scholars participate in supervised research both in the summer and during the academic year, and junior-year scholars learn about opportunities for summer research from our annual presentation on those opportunities.

One valuable opportunity is the Cal-Bridge summer research program, also known as CAMPARE (see figure 4). That program, which has been running since 2009, provides research opportunities for students from 21 CSU campuses and more than 30 community college partners to conduct summer research at one of 18 research sites around the country, including 7 UC campuses. Other scholars have obtained their own research placements with independent undergraduate research programs such as the Harvard–Smithsonian Astronomical Observatory, Northwestern CIERA, MIT Haystack Observatory, and

## The importance of mentoring

Cal-Bridge scholars highly value the mentoring they receive from the program. In our independent NSF grant evaluator's report, mentoring is listed as the most valuable part of the program, and scholars mention it twice as often as they mention the program's substantial financial support. Cal-Bridge graduates consistently say that they might not have reached their current levels of academic success without the Cal-Bridge program. The quotes below encapsulate the importance of Cal-Bridge mentoring for the students. (Photos courtesy of Cal-Bridge.)

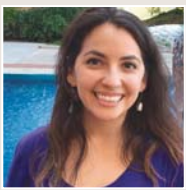
*The network of mentors and peers Cal-Bridge has helped me create has been invaluable in my pursuit of an astrophysics PhD! I now have an incredible support system of similarly underrepresented astro grad students and mentors who actively work to build a more inclusive community.*
—**Katy Rodriguez Wimberly** (BS, CSU Long Beach 2015; astronomy PhD candidate, UC Irvine; NSF Graduate Research Fellow)

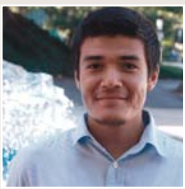
*If not for the immense support provided by all my mentors and peers in Cal-Bridge, I probably would have given up on my dream to go to grad school. They kept me moving forward.*
—**Luis Nuñez** (BS, Cal Poly Pomona 2018; astronomy PhD candidate, the Pennsylvania State University)

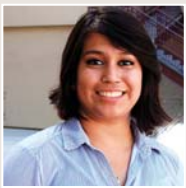
*Cal-Bridge opened up doors for me that led to great experiences which helped lead to where I am now. As a first-generation college student, they offered great resources and mentoring that helped guide me through school, internships, and graduate school.*
—**Becky Flores** (BS, CSU Northridge 2019; astronomy PhD candidate, Georgia State University)

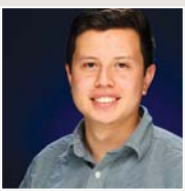
*"Help" is an understatement for what Cal-Bridge has done for me. Cal-Bridge prepared me academically and mentally to become a PhD candidate at UCI. Despite being an alumnus scholar, I still benefit from Cal-Bridge as I can connect with many other current and alumni scholars. My dream is to become a professor at a minority-serving institution. Cal-Bridge and CAMPARE have contributed to making this dream my career.*
—**Jeffrey Salazar** (BS, CSU San Bernardino 2018; astronomy PhD candidate, UC Irvine)

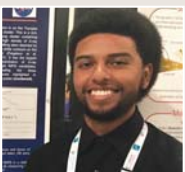
*Cal-Bridge is turning my dream of a PhD in Astrophysics into a tangible reality through required mentoring, personal academic support and substantial financial support. I would have never thought that a program like this was made for folks that resemble me and my background.*
—**Evan Nuñez** (BS, Cal Poly Pomona 2019; astronomy PhD candidate, Caltech; NSF Graduate Research Fellow)

the National Astronomy Consortium led by the National Radio Astronomy Observatory.

Working with UC faculty offers two distinct benefits to Cal-Bridge scholars. First, a summer research program gives the scholars a window into the type of research they might conduct if they attend that UC campus for graduate school. Second, a UC faculty member can get an in-depth look at a Cal-Bridge scholar in a research setting, so that they can speak to that scholar's capabilities in a letter of recommendation for graduate programs. The Cal-Bridge summer research program also acts as an additional recruiting mechanism for the main Cal-Bridge program by helping identify students who are likely candidates for it.

All Cal-Bridge scholars are expected to present the results of their research at regional and national conferences, such as the American Astronomical Society meeting or the APS March Meeting. Attending conferences and presenting research results are critical for students' professional development. A 2007 study of undergraduate research students noted that students who "became involved in the culture of research—attending conferences, mentoring other students, authoring journal papers, and so on—were the most likely to experience 'positive' outcomes," such as increased interest in pursuing a research career and increased likelihood of obtaining a PhD.[16]

In the fall of each year, Cal-Bridge hosts an annual research symposium together with our sister CAMPARE summer research program. Participants in both programs present their research results. Scholars' families are invited in order to enlist support for the students as they make their career choices, including the possible decision to pursue a PhD in physics or astronomy. Family support is critical for all students, but especially for URM and first-generation college students.[18]

## Outcomes

The Cal-Bridge program has already had a positive effect on the number of students from underrepresented groups pursuing physics and astronomy PhDs (see the box at left). The program has selected 59 scholars over the past five years. They include 34 Latinx, 7 African American, and 25 women students, with 15 of the women coming from underrepresented minority groups. Thus over 25% of Cal-Bridge scholars are women of color. Of the 59, 44 are first-generation college students. We were recently able to double our size, to 25 scholars per year, through a five-year, $5 million grant from NSF's S-STEM program. Growth is expected to continue with a long-term target of 35–40 scholars selected annually from across the state.

In the past four years, 27 of 33 (82%) Cal-Bridge scholars who graduated with a BS while in the program have begun PhD programs in physics or astronomy. Four others are enrolled in an APS Bridge Program or a master's degree program, and most are hoping to eventually earn a PhD. One scholar was accepted into a PhD program but chose to teach high school physics instead.

Of the 27 scholars who are in a PhD program, 10 are attending five UC programs: Davis, Irvine, Merced, Santa Barbara, and Santa Cruz. The other 17 scholars are attending 14 non-UC PhD programs across the country, including those at Caltech, Harvard, University of Maryland, Northwestern University, University of Pennsylvania, the Pennsylvania State University, and

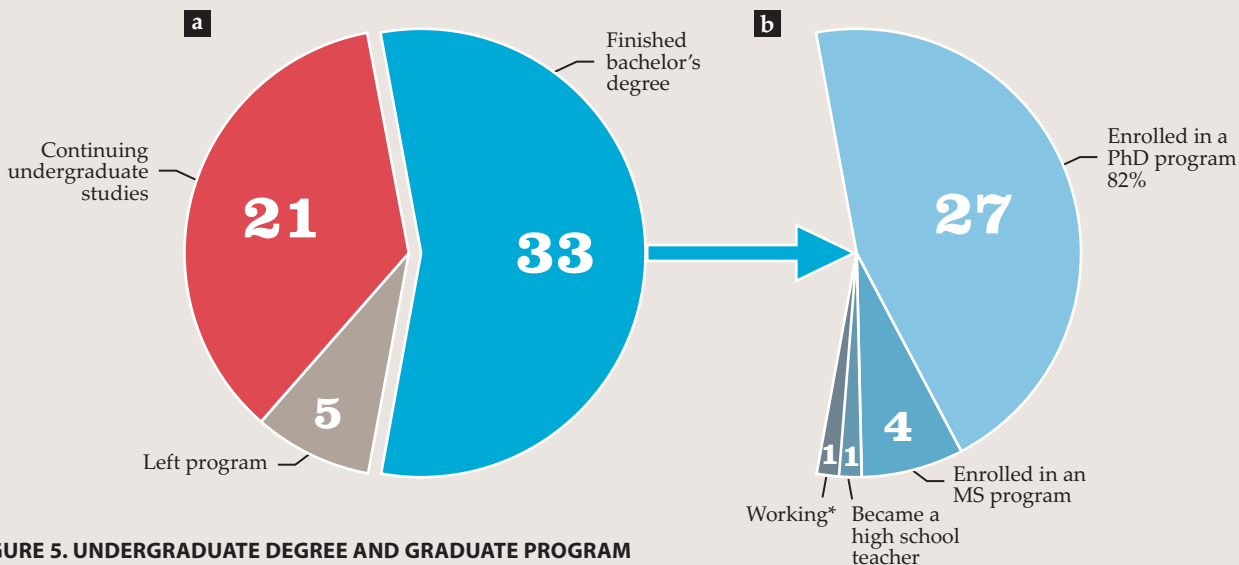

**FIGURE 5. UNDERGRADUATE DEGREE AND GRADUATE PROGRAM OUTCOMES FOR THE 59 CAL-BRIDGE SCHOLARS. (a)** Number of Cal-Bridge scholars who have completed their bachelor's degree, left the program, or continue their undergraduate studies. **(b)** Graduate program outcomes for Cal-Bridge students with bachelor's degrees.

*This is a scholar who decided to work while reapplying to PhD programs this fall.

University of Wyoming, among others. As mentioned previously, seven scholars have won NSF Graduate Research Fellowships, and four more received an honorable mention. Figure 5 shows the outcomes for the first five years of the program.

## Looking ahead

If Cal-Bridge expands to 40 scholars per year, we might expect 25–30 URM students from the program to pursue a PhD per year. That would increase the number of URM PhDs in physics and astronomy nationally by almost one-third.[1]

Administrators in both the CSU and UC systems are helping with plans to expand the program's reach in two ways: by expanding to other STEM fields, and by promoting emulation in other geographic regions. Program leadership has already reached out to faculty in computer science, mathematics, and engineering to talk about creating additional bridge programs in those fields, which have diversity issues similar to those of physics and astronomy.

In addition to expanding to other fields, we hope that this new model of an undergraduate–graduate PhD bridge, created as a network of regional universities, will be replicated in other parts of the country. Numerous regions of the US have high concentrations of minority-serving and Hispanic-serving institutions. Those regions include Texas and the southwestern US; the southeastern US, where many historically black colleges and universities are found; Florida, home to many Hispanic-serving institutions; and the New York metropolitan and Atlantic coast area. Cal-Bridge leadership is prepared to help any such regional partnership get off the ground with technical and other support, including sharing materials we developed and lessons learned.

To solve a problem as large and intractable as the lack of diversity in STEM in the US will require varied approaches and many programs. If the support for existing, successful programs continues and additional programs are created, we may achieve true equity in access and accomplishment in STEM fields.

*The author acknowledges the many faculty members at all three levels of the California public higher education system who have devoted countless volunteer hours to the success of the Cal-Bridge scholars. The author extends his deepest gratitude to the scholars of the program. Without their hard work and perseverance, the program would not be the success it has become.*

## REFERENCES

1. National Center for Science and Engineering Statistics, *2016 Doctorate Recipients from U.S. Universities*, NSF (March 2018).
2. National Academy of Sciences, National Academy of Engineering, and Institute of Medicine, *Expanding Underrepresented Minority Participation: America's Science and Technology Talent at the Crossroads*, National Academies Press (2011), p. 36.
3. US Department of Labor, *Current Employment Statistics*, Bureau of Labor Statistics (2017), Table A-1, Employment status of the civilian population by sex and age.
4. J. R. Posselt, *Inside Graduate Admissions: Merit, Diversity, and Faculty Gatekeeping*, Harvard U. Press (2016).
5. S. Hurtado et al., *J. Soc. Issues* **67**, 553 (2011).
6. A. M. Porter, R. Ivie, *Women in Physics and Astronomy, 2019*, AIP Statistical Research Center (2019); American Institute of Physics, "Minority Issues" (2019).
7. K. G. Stassun, A. Burger, S. Edwards Lange, *J. Geosci. Educ.* **58**, 135 (2010).
8. K. G. Stassun et al., *Am. J. Phys.* **79**, 374 (2011).
9. S. Edwards Lange, "The master's degree: A critical transition in STEM doctoral education," PhD thesis, U. Washington (2006).
10. National Center for Science and Engineering Statistics, *Women, Minorities, and Persons with Disabilities in Science and Engineering*, NSF rep. 19-304, NSF (2019).
11. Ref. 2, p. 12.
12. S. E. Cross, N. V. Vick, *Pers. Soc. Psychol. Bull.* **27**, 820 (2001).
13. E. Armstrong, K. Thompson, *J. Women Minor. Sci. Eng.* **9**, 159 (2003).
14. J. R. Posselt, *Rev. Higher Educ.* **41**, 497 (2018).
15. S. Hurtado et al., *New Dir. Inst. Res.* **2010**(148), 5 (2010).
16. S. H. Russell, M. P. Hancock, J. McCullough, *Science* **316**, 548 (2007).
17. M. K. Eagan Jr et al., *Am. Educ. Res. J.* **50**, 683 (2013).
18. S. Slovacek et al., *J. Res. Sci. Teach.* **49**, 199 (2012).

PT